\documentclass[%
reprint,
superscriptaddress,
amsmath,amssymb,
aps,
prb,
floatfix
]{revtex4-2}

\usepackage{multirow}
\usepackage{graphicx}
\usepackage{dcolumn}
\usepackage{bm}
\usepackage[version=3]{mhchem}
\usepackage{nicefrac}

\usepackage{color}
\usepackage{dcolumn}
\usepackage{amssymb}
\usepackage{url}
\usepackage{hyperref}
\usepackage{xfrac}
\usepackage{tabularx}
\usepackage{xspace}
\usepackage{amsmath}
\usepackage[normalem]{ulem}
\usepackage{mathtools}
\usepackage{physics}
\usepackage{dsfont}

\def\be{\begin{equation}}
\def\ee{\end{equation}}
\def\bea{\begin{eqnarray}}
\def\eea{\end{eqnarray}}
\def\ba{\begin{array}}
\def\ea{\end{array}}

\begin{document}
\title{Magnetization of Antiferromagnetic Cactus Graph Model with Exact Dimer Ground State}
\author{Pratyay Ghosh}
\email{pratyay.ghosh@uni-wuerzburg.de}
\affiliation{Institut f\"ur Theoretische Physik und Astrophysik and W\"urzburg-Dresden Cluster of Excellence ct.qmat, Universit\"at W\"urzburg,
Am Hubland Campus S\"ud, W\"urzburg 97074, Germany}


\begin{abstract}
We introduce and explore the magnetization behavior in a quantum spin system on a cactus graph, deemed the Cactus Graph Model (CGM), featuring an exact dimer singlet ground state. We analyze the singlet-triplet gap, interactions among excited triplets, and correlated hopping under an external magnetic field using a strong coupling expansion. Employing an effective hard-core boson representation, we unveil density wave magnetization plateaus and supersolid phases, revealing the intricate interplay between interactions and hopping dynamics. Furthermore, we conjecture that correlated hopping gives rise to multi-triplet bound states, and potentially engendering to the stabilization of additional low-lying magnetization plateaus.
\end{abstract}

\maketitle
\section{Introduction} In the field of condensed matter physics, the Cayley tree (CT)~\cite{Cayley1878} and Bethe lattice (BL)~\cite{Bethe1935} are fundamental graph structures used to explore spin systems and their complex statics and dynamics. Despite the shift in contemporary graph theory towards realistic random graphs~\cite{Erdoes1959} and complex networks~\cite{Albert2002,Dorogovtsev2002,Dorogovtsev2008}, the enduring significance of BL and CT lies in their precise tree-like architecture devoid of cycles~\cite{Ostilli2012}. This characteristic positions BL and CT as instructive models for conducting exact calculations~\cite{Baxter1982}, providing a simple yet insightful comprehension of interactions and cooperative behaviors, and capturing emergent behaviors, critical phenomena, and phase transitions in spin systems~\cite{Diepbook,frustrationbook}.

While exploring spin systems, the investigation of exactly solvable models presents a captivating avenue~\cite{Majumdar1969,Shastry1981,Ghosh2022,Ghosh2023a,SSM_Review}. A typical example is the famous Shastry-Sutherland model~\cite{Shastry1981}, a frustrated spin system on a lattice of orthogonal dimers, which exhibits an exact dimer singlet ground state. The exact solvability of this model offers profound insights into exotic phases, quantum entanglement, and novel phase transitions~\cite{SSM_Review,Koga2000,Knetter2000,Totsuka2001,Misguich2001,Corboz2014,Lee2019,Shi2022}. Thus, the fusion of the BL's analytically tractable nature with the intricate details of exact dimer models can create a fertile foundation for comprehending the diverse behaviors of quantum spin systems.

\begin{figure}
\includegraphics[width=\columnwidth]{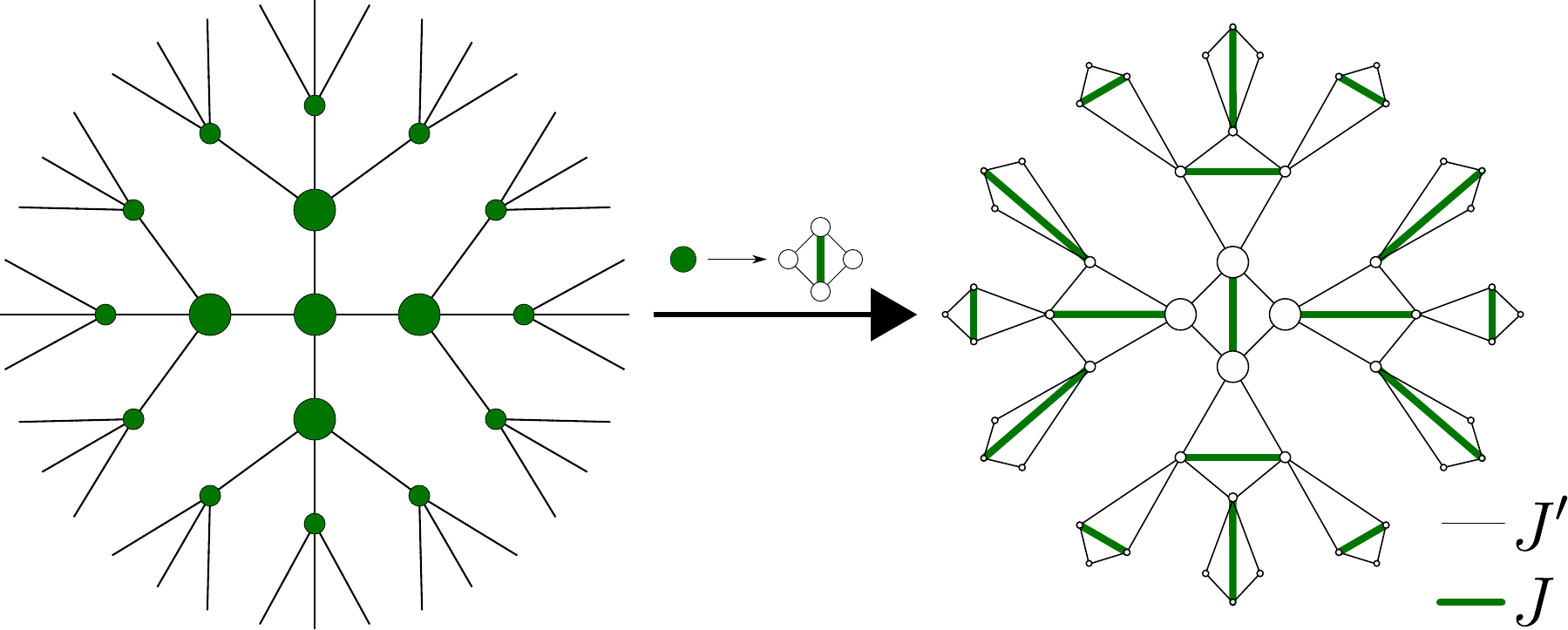}
\caption{The cactus graph model (CGM): a graph structure constructed from a Bethe lattice of degree $z=4$. The construction involves generating a graph of quadrangles with shared corners. The depicted diagonals represent the dimer bonds of the model. The singlets residing on the dimer bonds represent an eigenstate of~\eqref{eq-hamil}.} \label{fig-tree}
\end{figure}

\begin{figure*}
\includegraphics[width=0.9\textwidth]{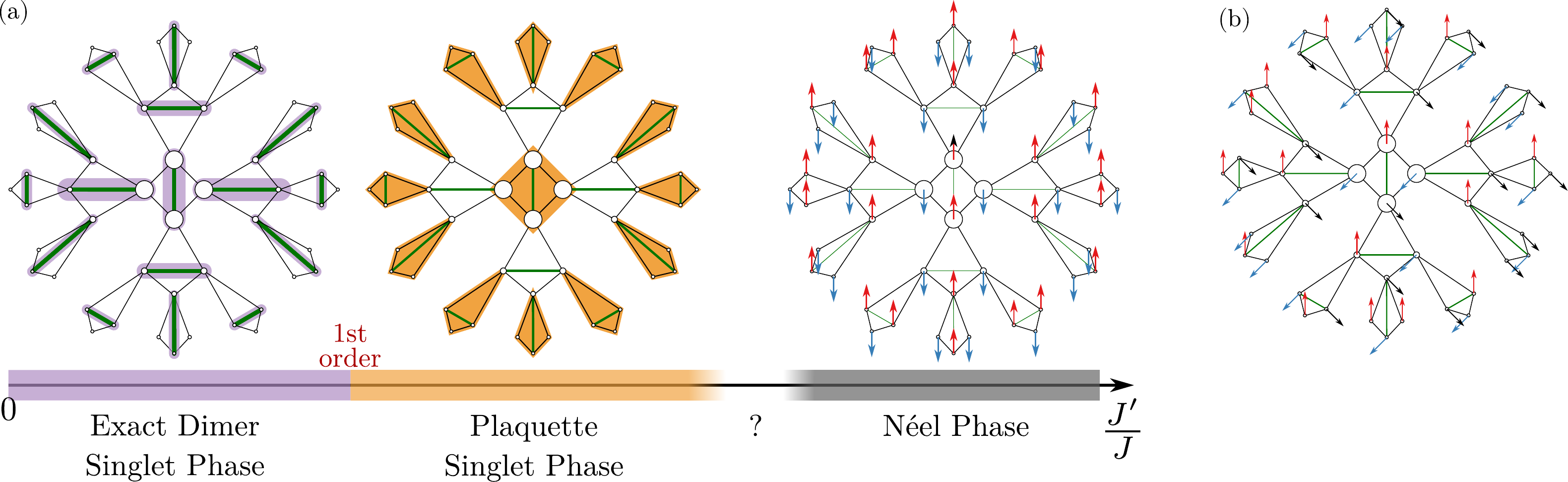}
\caption{(a) A speculative quantum phase diagram of~\eqref{eq-hamil}. The schematics illustrate three states: the exact dimer singlet state, the plaquette singlet state, and the N\'eel phase. The purple ellipses represent exact dimer singlets, and the orange quadrangles represents plaquette singlets. Like the other exact dimer models~\cite{Shastry1981,Ghosh2022,Ghosh2023a}, the transition out of the dimer singlet phase is likely to be first order. The phase transition from the plaquette singlet phase to the N\'eel phase may occur at a deconfined quantum critical point~\cite{Lee2019} or they could be separated by an intermediate spin liquid phase~\cite{Shi2022}. (b) For the CGM with classical spins, i.e. $S\to\infty$, there exists an incommensurate spiral state for $J'<J$, which is different from the N\'eel state. This state is highly degenerate and incommensurate except for some particular ratios of $J'/J$ where $\arccos(J'/J)$ is a rational fraction of $\pi$. One such configuration for $J'=J/2$ is shown.} \label{fig-qpd}
\end{figure*}

In this article, we start by constructing a quantum spin model on a cactus graph (CG) that admits an exact dimer singlet eigenstate. We commence with a BL of degree $z=4$ and follow the protocol in~\cite{Ghosh2023a} to generate a CG comprising quadrangles with shared corners. This involves decorating all edges of the BL with new vertices, which are the physical sites hosting spin-$1/2$'s, and then employing a Y-$\Delta$-type transformation (Fig.~\ref{fig-tree}) to eliminate the original vertices and establish pairwise connections between the spins. These newly formed edges act as inter-dimer bonds. At this stage, the engineered graph structure (the medial graph of the original BL) is a CG. Next, we connect one diagonal of each quadrangle (indicated by the thick green lines), avoiding site sharing between any two dimers. These diagonals constitute our dimer bonds (depicted in Fig.~\ref{fig-tree}). The model Hamiltonian is given by
\be\label{eq-hamil}
\hat{H}=\hat{H}_0+\hat{H'}
\ee
with
\bea
\hat{H}_0&=&J\sum_{\langle ij\rangle}\mathbf{\hat{S}}_i\cdot\mathbf{\hat{S}}_j\\
\hat{H'}&=&J'\sum_{\langle kl\rangle}\mathbf{\hat{S}}_k\cdot\mathbf{\hat{S}}_l.
\eea
Here, $\mathbf{\hat{S}}_i$ denotes a spin operator acting on a spin-$1/2$ degree of freedom at site $i$. There are two separate summations $ij$, and $kl$ over the thick green and black bonds, respectively (Fig. \ref{fig-tree}). We deem this model as the cactus graph model (CGM). It can be analytically shown that for antiferromagnetic spin exchange interactions ($J, J'>0$), the model has an exact ground state, i.e. a product state of singlets on the dimer-bonds, for $J'\le J/2$ [see left most panel of Fig.~\ref{fig-qpd} (a)]. The demonstration is straightforward. The product state of singlets, i.e.
\be|\psi\rangle=\otimes\prod_{\langle ij\rangle}|s\rangle_{ij},\ee
is the ground state of the first term in~\eqref{eq-hamil}. Here, $|s\rangle_{ij}$ denotes a singlet constituted of the spins at site $i$ and $j$. Now, the matrix element of the second term in~\eqref{eq-hamil} nullifies for any adjacent pair of $J'$ bonds due to the singlet's odd parity concerning the reflection that interchanges the two spins, leading to $\hat{H'}|\psi\rangle=0$. This makes the dimer state an exact eigenstate of~\eqref{eq-hamil} with energy $e_0=-\frac{3}{8}JN$ ($N$ is the total number of spins). This state serves as the ground state for when $J'$ is sufficiently smaller than $J$ -- an analysis based on variational principle finds it to be $J'\le J/2$ (see Refs.~\cite{Shastry1981,Ghosh2022,Ghosh2023a} for further details). This singlet state can also remain as the ground state for $J'>J/2$, only it cannot be shown analytically~\cite{Koga2000,Chung2001,Corboz2013,Lee2019,Yang2022,Farnell2011,Ghosh2022,Ghosh2023a}. 

As the system is bipartite for $J=0$, it accommodates a N\'eel-type state near that limit [see Fig.~\ref{fig-qpd} (a)]. Note that the classical phase diagram ($S\to\infty$) of CGM also features a spiral phase for $J'<J$ akin to SSM. The spiral state where the angle between the spin across the $J'$ bonds are $\pi-\phi$ and the same between the spins across the $J$ bonds is $2\phi$, where
\be
\phi=\arccos(J'/J).
\ee
For $S=1/2$ SSM, this classical highly degenerate incommensurate spiral phase translates into a plaquette singlet phase separating the exact dimer singlet phase and the N\'eel phase~\cite{Shastry1981}. Such a plaquette singlet phase can also be predicted for CGM due to the same reason [see Fig.~\ref{fig-qpd} (b)], however, the details of the state are slightly different from the one found in SSM~\cite{Koga2000,Corboz2013,Lee2019,Yang2022}. As the plaquette singlet and the N\'eel state each break distinct lattice and spin-rotation symmetries, a second-order transition between them is beyond the conventional Landau-Ginzburg paradigm, hinting at a deconfined quantum critical point~\cite{deconfined,Lee2019}. Alternatively, there might exist a quantum spin-liquid sandwiched between the two phases~\cite{Chandra1994,Shi2022}.

In this article, we do not study the phase diagram of the model but focus on another interesting avenue of a model with an exact dimer ground state with a Zeeman term, $-B\sum_{i}S_i^z$, with the magnetic field, $B$, of the order of the dimer singlet-triplet gap, i.e., $B\sim J$. After a perturbative expansion in $J'$, we perform a hard-core boson (HCB) projection onto a two-dimensional basis, comprising the dimer singlet and the field-aligned triplet. The effective HCB model exhibits an appealing system devoid of single-particle dynamics and dictated by two-body interactions and correlated hoppings~\cite{Miyahara1999,Miyahara2000,Ghosh2023}. As a result, the system can break translation symmetry to form density waves, correlated hopping may lead to bound states~\cite{Knetter2000,Hard-Boson-SSM,Totsuka2001,Manmana_2011,Corboz2014,Ghosh2023}, and supersolid non-plateau states~\cite{Hard-Boson-SSM,Wierschem2018,Shi2022} where density wave and superfluidity coexist.

\section{Singlet-triplet gap}
In the exact dimer phase, i.e. $J'\lesssim J/2$, the ground state of~\eqref{eq-hamil} is a product state of spin singlets on all dimer bonds. Isolated dimer eigenstates comprise a singlet state $|s\rangle$ and three triplet states $|t_{0,\pm 1}\rangle$ (the subscript denotes the total $S^z$ quantum number). The second term in~\eqref{eq-hamil}, i.e. the inter-dimer interactions, is incorporated perturbatively on the basis of dimer eigenstates~\cite{Sakurai2020}. The singlet-triplet gap then reads:
\be\label{eq-gap}
\Delta = J - B - \frac{J'^2}{J} - \frac{J'^3}{2J^2} - \frac{J'^4}{8J^3} + \mathcal{O}\left(J'^5\right).
\ee

For exact dimer models, such as the SSM~\cite{Shastry1981}, Maple leaf model (MLM)~\cite{Ghosh2022}, or the current Cactus Graph model (CGM), the singlets' parity and the matrix elements of $J'$ terms establish limitations on the hopping dynamics of the triplets. Specifically, a triplet can move solely to two out of its four adjacent dimers [see Fig.~\ref{fig-int-hop} (a)] as
\be
\hat{h}'_{\alpha\beta}\ket{s}_\alpha\ket{t_{\pm1,0}}_\beta=0
\ee
where $\hat{h}'_{\alpha,\beta}$ is the $J'$ terms acting between the dimers $\alpha$ and $\beta$. In the process, dictated by the matrix elements
\bea
\hat{h}'_{\beta\gamma}\ket{t_{\pm1}}_\beta\ket{s}_\gamma&=&\pm\frac{J'}{2}\left(\ket{t_{\pm1}}_\beta\ket{t_{0}}_\gamma-\ket{t_{0}}_\beta\ket{t_{\pm1}}_\gamma\right)\\
\hat{h}'_{\beta\gamma}\ket{t_{0}}_\beta\ket{s}_\gamma&=&\frac{J'}{2}\left(\ket{t_{1}}_\beta\ket{t_{-1}}_\gamma-\ket{t_{-1}}_\beta\ket{t_{1}}_\gamma\right),
\eea
it leaves another triplet behind~\cite{Miyahara1999,Ghosh2023}. For MLM and SSM, the underlying lattice of dimers contains cycles of girth three and four, respectively, and, hence, single triplet hoppings start in fourth- and sixth-order in perturbation, respectively~\cite{Miyahara1999,Miyahara2000,Ghosh2023}. For CGM, however, the graph of dimers is a BL, i.e. cycle-free, and thus, the single triplet hopping is entirely suppressed. The same is also true for a spin chain of linked tetrahedra~\cite{Gelfand-1991}. As the single particle dynamics lowers the spin-gap, the $\Delta$ for CGM provides an upper bound to the singlet-triplet gap for all exact dimer models.

\section{Interactions between excited triplets.}
\begin{figure}
\includegraphics[width=0.92\columnwidth]{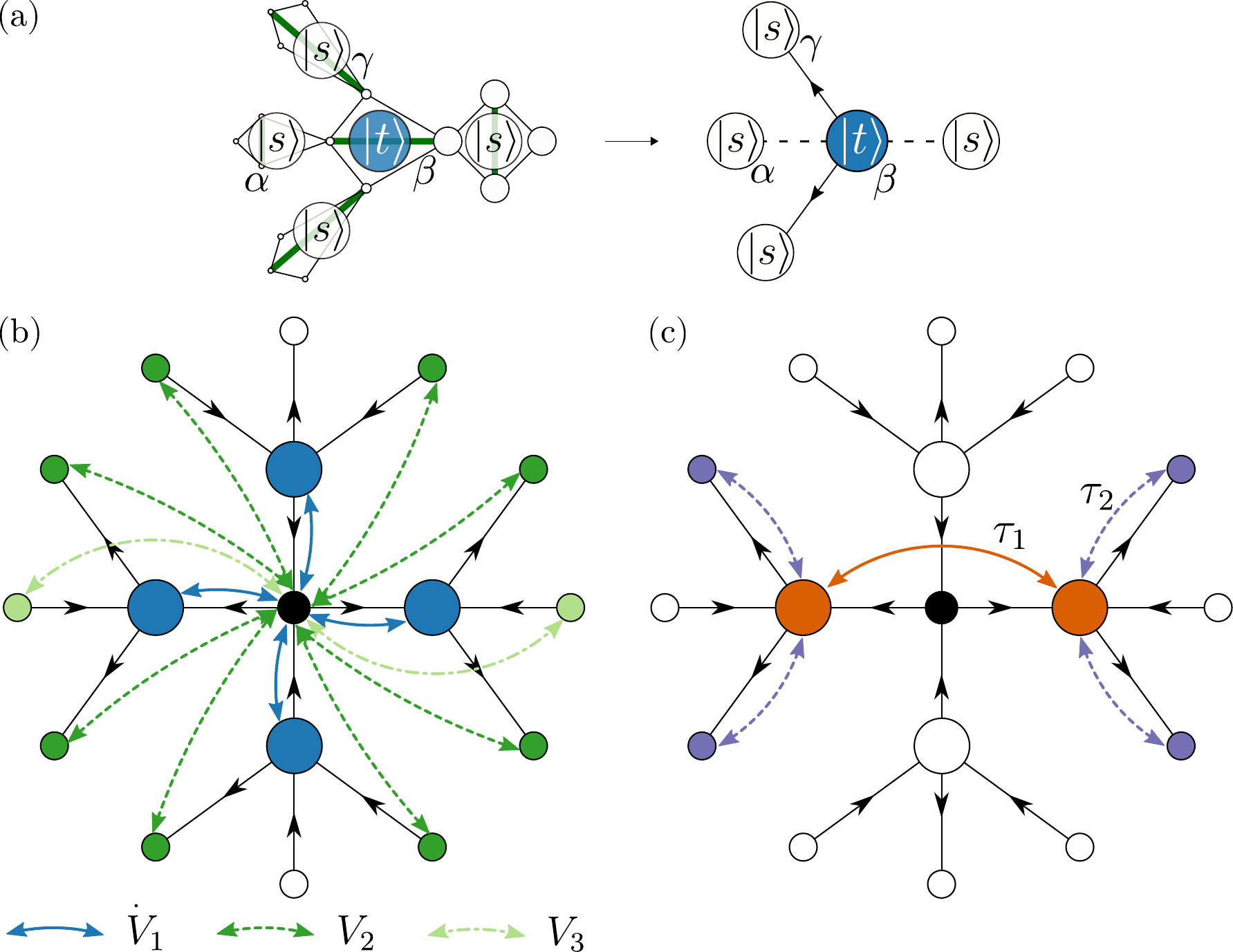}
\caption{(a) The singlet's parity and the matrix elements of the $J'$ terms~\cite{Ghosh2023} restrict the hopping of the triplets only to two of its four adjacent dimers. The possible hopping directions are indicated using arrows on the edges. (b) The interactions between pairs of triplets (reference triplet in black) in the CGM. The interactions are determined within third-order perturbation theory. (c) Correlated hopping in CGM, which is the movement of one triplet (filled in purple or orange) assisted by the other (filled in black).} \label{fig-int-hop}
\end{figure}

We, now, discuss the interactions between the excited $S^z=1$ triplets ($|t_1\rangle$) determined within the perturbation theory. We estimate the energy needed to generate two triplet excitations from the ground state. The computed energy contains two components: (i) the spin-gap energy required for two triplet excitations ($2\Delta$) including the self-energy corrections in~\eqref{eq-gap}, and (ii) the interaction energy between the excited triplets ($V_n$'s). Fig.~\ref{fig-int-hop} (b) provides a visual depiction of the interactions occurring between two triplets within third-order perturbation theory. The $V_n$'s are given by,
\bea
V_1&=&\frac{J'}{2}+\frac{J'^2}{2J}-\frac{J'^3}{8J^2}+\mathcal{O}\left(J'^4\right)\\
V_2&=&\frac{J'^3}{8J^2}+\mathcal{O}\left(J'^4\right)\\
V_3&=&\frac{J'^2}{2J}+\frac{3J'^3}{4J^2}+\mathcal{O}\left(J'^4\right).
\eea
Note that the $V_n$'s are closely related to the two triplet interactions found in SSM. The interaction energies display strong directional dependence and lack monotonicity. These characteristics are rooted in the geometry of the model. The triplet-triplet interactions extend beyond third-order, yet higher-order calculations are analytically intricate. Here advanced numerical techniques~\cite{Dorier2008,Corboz2014, Liu2014} will be advantageous and are to be pursued in the future.

\section{Correlated Hopping}
A notable feature inherent to the dimer model lies in its ability to facilitate two-triplet hopping, termed correlated hopping, which involves the motion of two triplets in a concerted manner [see Fig.~\ref{fig-int-hop} (c)]. Such a process defies categorization as either single-triplet hopping or the creation-annihilation of a triplet pair. This correlated hopping involving two triplets starts from second-order perturbations. The hopping amplitudes (which are the same for SSM) are given by
\bea
\tau_1&=&\frac{J'^2}{4J}+\frac{3J'^3}{8J^2}+\mathcal{O}\left(J'^4\right)\\
\tau_2&=&\frac{J'^2}{4J}+\frac{5J'^3}{16J^2}+\mathcal{O}\left(J'^4\right).
\eea
Given the suppressed single triplet dynamics, correlated hopping assumes a pivotal role. That has likewise been predicted in the SSM and MLM, where correlated hopping leads to the formation of multi-triplet bound states~\cite{Knetter2000,Hard-Boson-SSM,Totsuka2001,Manmana_2011,Corboz2014,Ghosh2023}.

\begin{figure*}
\includegraphics[width=0.8\textwidth]{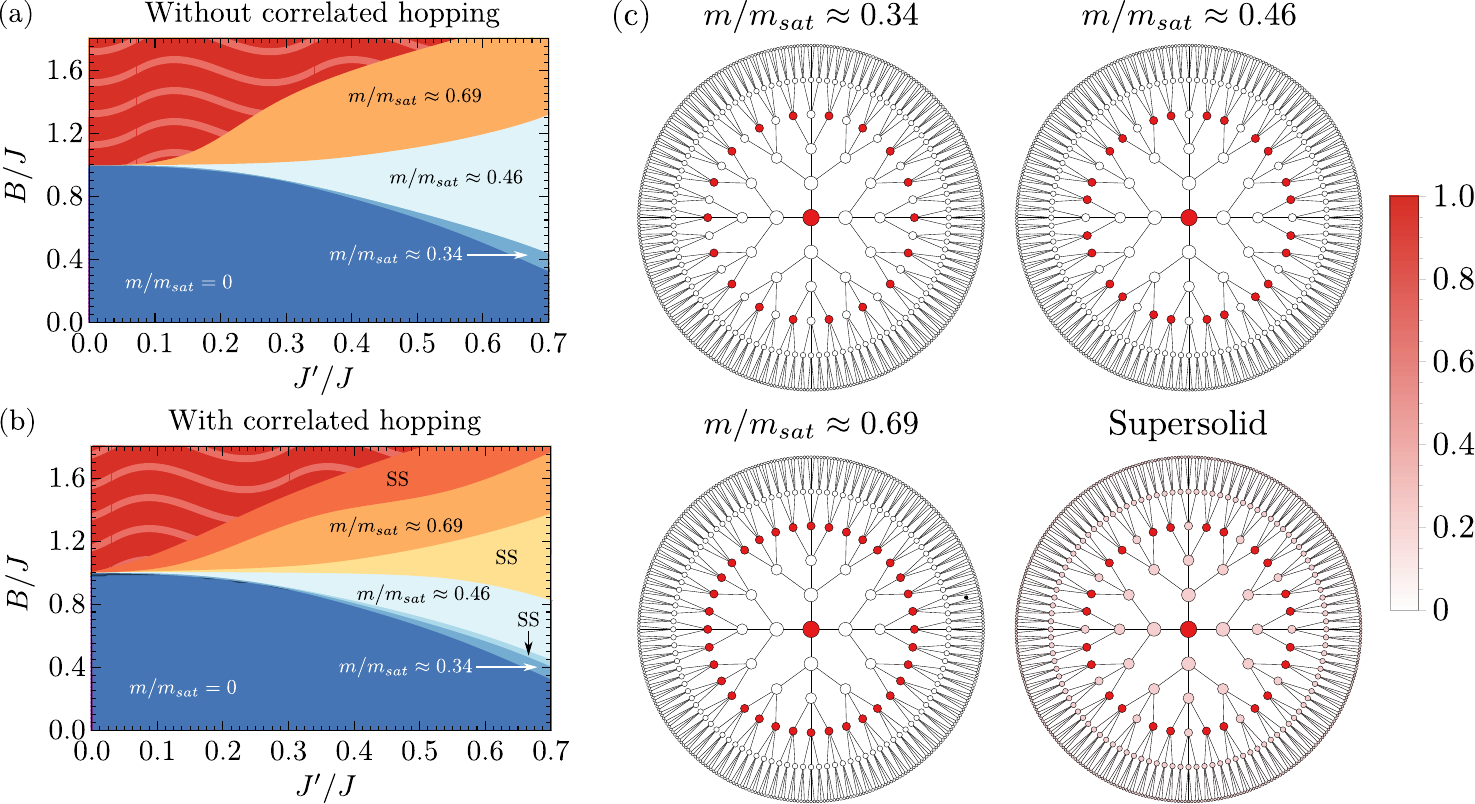}
\caption{Magnetization phase diagram of the spin Hamiltonian achieved via Matsubara-Matsuda transformation of the effective hard-core boson Hamiltonian~\eqref{eq-HC_bosons} (a) without and (b) with correlated hopping for a $1457$-site Cayley tree. The diagrams highlight magnetization plateaus at specific fractional values. Correlated hopping introduces additional supersolid (SS) phases that influence the stability of density wave magnetization plateaus. The wavy-patterned regions are not accessible within our current calculations. (c) The density waves and a representative supersolid state (central $485$ sites are shown). The color represents the local densities of the hard-core bosons. The density wave phase exhibits a regular spatial pattern, indicating a stable arrangement of particles. The supersolid phase demonstrates the coexistence of density wave behavior and superfluidity.} \label{fig-CPD}
\end{figure*}

\section{Effective hard-core boson model}
We, now, proceed to establish a mapping of the initial Heisenberg model onto a HCB model on the lattice of the dimers, i.e. a $z=4$ BL. For $B>0$, the system's lowest energy states with magnetization $m\ge 0$ consist of $|s\rangle$ and $|t_1\rangle$. These states are projected as low-energy physical degrees of freedom. By treating $|t_1\rangle$ as a HCB and $|s\rangle$ as a vacancy, we execute a projection, yielding an effective HCB model containing two-particle interactions and correlated hopping up to third order in perturbation theory. In the perturbative treatment, $|t_1\rangle$ triplets are treated as physical magnetic entities, and other dimer states — $|t_0\rangle$ and $|t_{-1}\rangle$ — serve as intermediate virtual states within the perturbation theory. The resultant effective Hamiltonian can be expressed in the general form:
\bea\label{eq-HC_bosons}
\hat{H}_\text{eff}=&&\Delta\sum_{r}\hat{n}_{r}+\sum_nV_{n}\sum_{(r,r')_n}\hat{n}_{r}\hat{n}_{r'}\nonumber\\
&&+\sum_m\tau_m \sum_{\left[r,r',r''\right]_m}\hat{n}_{r}\left[\hat{b}_{r'}^\dagger \hat{b}_{r''}+h.c.\right].
\eea
The resulting Hamiltonian contains onsite potential, mutual repulsion, and correlated hopping. The index $r$ runs over the sites of the effective BL of dimer bonds, $(r,r')_n$ sums span the pair of sites that interact via $V_n$ [see Fig.~\ref{fig-int-hop} (b)], and $\left[r,r',r''\right]_m$ sum represents the correlated hoppings with amplitude $\tau_m$, as shown in Fig.~\ref{fig-int-hop} (c). The operator $\hat{b}_r^\dagger$ ($\hat{b}_r$) creates (annihilates) a $|t_1\rangle$ particle at bond $r$, and $\hat{n}_r=\hat{b}_r^\dagger \hat{b}_r$ represents the particle number operator.

To study the Hamiltonian of HCBs~\eqref{eq-HC_bosons}, we take a classical approach, whereby HCB operators are mapped to spin-$1/2$ operators using the Matsubara-Matsuda transformation~\cite{Matsubara1956,Mila2008,Schmidt2008}:
\be
\hat{n}_{r}=\frac{1}{2}-\hat{T}_{r}^z\text{,\ \ }\hat{b}_{r}=\hat{T}_{r}^+\text{,\ \ }\hat{b}_{r}^\dagger=\hat{T}_{r}^-.
\ee
Note that, this transformation is exact. Following this transformation,~\eqref{eq-HC_bosons}'s onsite potential and repulsion terms manifest as Ising terms and a longitudinal magnetic field, while correlated hopping introduces $XX$-type terms, along with additional three-spin interactions.

The diverse phases of the bosonic Hamiltonian, $\hat{H}_\text{eff}$, can be read off from the spin configuration -- an empty site is represented as a spin-up state, while an occupied site corresponds to a spin-down state. Therefore, the density $\hat{n}$ is replaced by the magnetization per site $m=\frac{1}{2}-\hat{n}$. Note that, the occupation can only happen when the spin gap $\Delta$ is closed by the physical magnetic field, $B$. Nonetheless, the robust repulsive interactions prevent all singlets from transitioning to triplets upon gap closure. Due to the interplay between the energy gains from triplet excitations and the repulsive energy, for a given $B$, triplets would adopt a distinct superstructure (density wave) to optimize energy. This density wave, with constant magnetization, remains a stable configuration within a finite range of $B$. It demonstrates long-range correlations and displays a spin gap. 

For other plausible phases of the bosonic Hamiltonian, such as the superfluids and supersolids, $\langle \hat{b}_r\rangle=\langle\hat{T}_{r}^+\rangle$ serves as the order parameter. A superfluid state is a ferromagnetic state with all the spins having a non-zero projection onto the $xy$-plane. A supersolid phase, which is a superfluid state on the background of a density wave, becomes a long-range order with a fraction of spins pointing down and the rest having a non-zero projection onto the $xy$-plane.

In our approach, we adopt a classical treatment of the effective spin Hamiltonian. This involves assuming that the spin operators, $\mathbf{\hat{T}}_r\equiv\{\hat{T}^z,\hat{T}^+,\hat{T}^-\}$, can be approximated as classical vectors of length $1/2$. The ground state is then determined by arranging the spins in a configuration that minimizes the classical Hamiltonian. This minimization process is conducted numerically using the iterative minimization technique~\cite{Sklan2013} on a CT comprising $1457$ sites. The approach involves initialization of the system with a random spin configuration, then iteratively selecting one vector-spin, $\vec{T}_r$, at a time and aligning it antiparallel to its local field, i.e. a directive given by $\partial \hat{H}_\text{eff}/\partial \vec{T}_r$. This minimizes the overall energy of the system. In the course of each iteration, we ensure that, on average, each spin experiences a single update. We conduct extensive $10^7$ sweeps for each set of parameters, and convergence is recognized when the total energy difference per iteration falls below the threshold of $10^{-4}$ in the unit $J$.

The resulting phase diagrams, both in the absence and incorporation of correlated hopping, are depicted in Fig.~\ref{fig-CPD} (a) and (b), respectively. Notably, both phase diagrams exhibit magnetization plateaus at $m/m_{\text{sat}}\approx0.34$, $0.46$, and $0.69$. Note that the finiteness of CT can induce severe boundary effects on the plateau states. Therefore, in a BL, the magnetization might differ from the ones presented by the CT. We, however, do not embark on that study in detail -- our purpose here is to demonstrate the essential physics that the CGM has to offer.

When considering the phase diagram in the context of correlated hopping, additional supersolid phases emerge in the vicinity of criticality. The supersolid phases can be understood in the following manner: as an external field destabilizes a plateau state, the system retains the density wave characteristic of the preceding phase, while also accommodating extra bosons. These bosons, when assisted by the bosonic density wave, achieve superfluid-like mobility in the lattice through correlated hopping. As a consequence, a supersolid phase, a combination of density wave and the superfluid~\cite{Hard-Boson-SSM,Wierschem2018,Shi2022}. Within this phase, the magnetization increases monotonically with increasing field until the next plateau state is reached. We do not explore the magnetization beyond the $0.69$ plateau, as from there one encounters a dense population of HCBs, necessitating the consideration of three-particle interactions to thoroughly explore higher magnetization plateaus. The various bosonic density wave patterns and a representative supersolid structure are depicted in Fig.~\ref{fig-CPD} (c).

\begin{figure*}
\includegraphics[width=0.8\textwidth]{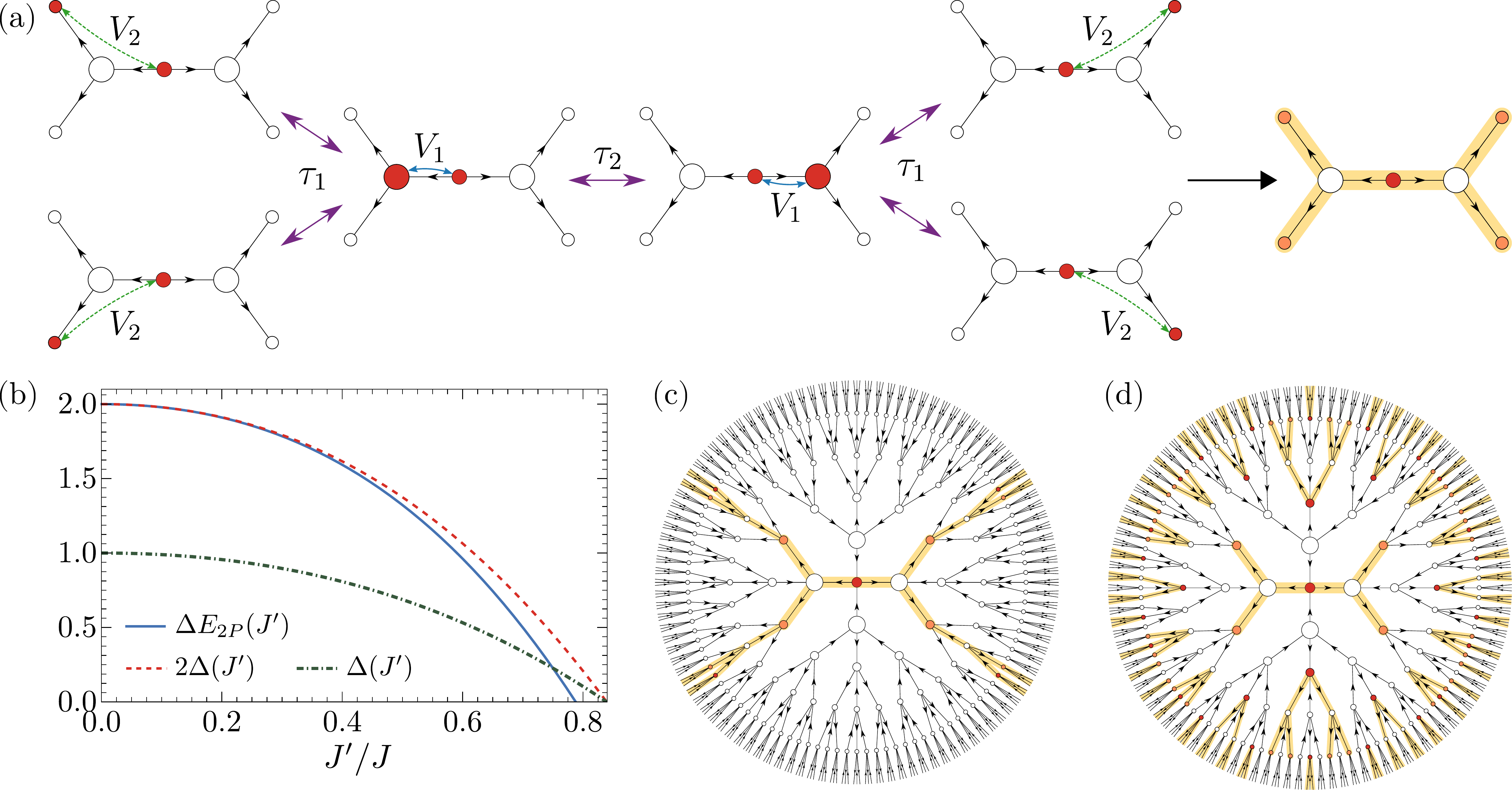}
\caption{(a) Correlated hopping facilitates the motion of two excited triplets (filled in red), leading to two-triplet bound states indicated in light-yellow. (b) The dot-dashed line represents $\Delta$, while the dashed line represents the energy associated with two noninteracting dimer triplets ($2\Delta$), and the solid line corresponds to the energy of a bound state of two triplets, $\Delta E_{2P}$, (in the unit of $J$) as a function of $J'/J$. Example of (c) an extended multi-triplet bound state (d) a crystal of two-triplet bound states.} \label{fig-corr-hopp}
\end{figure*}

\section{Triplet bound states}
While two excited triplets are distantly separated, each triplet remains localized. However, when they are adjacent to each other, the phenomenon of correlated hopping facilitates the coherent movement of the two triplets and the system gains kinetic energy. When two triplets are placed on nearest-neighbor dimers, their interaction, linear in $J'$, is typically the dominant factor. They can, however, utilize correlated hopping ($\propto J'^2$) to separate from each other. This separation reduces their repulsive energy cost to $V_2$, which is $\propto J'^3$. As a result, they effectively form a two-particle bound state. See Fig.~\ref{fig-corr-hopp} for a visual representation of the relative motion of two triplets. By diagonalizing the two-particle hopping matrix
\be
2\Delta\times\mathbf{I}_6+\begin{bmatrix}
V_2 & 0 & \tau_2 & 0 & 0 & 0 \\
0 & V_2 & -\tau_2 & 0 & 0 & 0 \\
\tau_2 & -\tau_2 & V_1 & \tau_1 &0 &0 \\
$0$ &0 & \tau_1 & V_1 & -\tau_2 & \tau_2 \\
$0$ &0 &0 & -\tau_2 & V_2 &0 \\
$0$ &0 &0 & \tau_2 &0 & V_2
\end{bmatrix},
\ee
we evaluate the energies of a two-triplet excitation to identify a preferred two-triplet bound state with energy
\bea
\Delta E_{2P}=2\Delta&+&\frac{1}{2}\left[\left(V_1+V_2-\tau_1\right)\vphantom{\sqrt{\left(V_1+V_2-\tau_1\right)^2+8\tau_2^2}}\right.\nonumber\\
&-&\left.\sqrt{\left(V_1+V_2-\tau_1\right)^2+8\tau_2^2}\right]
\eea
that occupies the indicated light-yellow region in Fig.~\ref{fig-corr-hopp} (a). This bound state primarily comprises a linear combination of two-triplet states that interact via $V_2$ and has a kinetic energy gain of $-\frac{J'^3}{8J^2}+O(J'^4)$ from the two-particle threshold. Note that, in SSM~\cite{Hard-Boson-SSM,Totsuka2001}, the bound state is formed on a cycle of length four. Therefore, the two-particle bound state in CGM is different from the one in SSM~\cite{Hard-Boson-SSM,Totsuka2001} due to the cycle-free nature of the BL of the dimers. As a consequence of the two-particle bound states possessing lower energy than two distinct triplets, the gap associated with these bound states decreases more rapidly compared to the singlet-triplet gap when $J'$ is increased [see Fig.~\ref{fig-corr-hopp} (b)]. Consequently, this leads to the precedence of bound states undergoing condensation before isolated individual triplets for $0.75\lesssim J'/J\lesssim 0.79$.

In the case of SSM and MLM, it has been predicted that $S^z=2$ triplet bound states cause the emergence of extra magnetization plateaus close to $m=0$ through the creation of bound state crystals~\cite{Corboz2014,Schneider2016}. A similar phenomenon can be anticipated for the CGM also. It must also be mentioned that the formation of a bound state is not limited to two triplets, but the system can produce extended bound states with large $S^z$. Fig.~\ref{fig-corr-hopp} (c) shows an example of such extended bound states. In Fig.~\ref{fig-corr-hopp} (d), we predict a crystal of two-triplet bound states on the BL, which is a supersolid state giving rise to low-magnetization plateau states.

We expect that these bound states no longer persist as the magnetization increases. More precisely, single-triplet excitations can gain mobility via correlated hopping at high magnetizations, and they can amass higher kinetic energy compared to bound states. This implies a transition of elementary particles from bound states at low magnetizations to single-triplet excitations at high magnetization.

\section{Conclusions and Outlook}
In this study, we have introduced and undertaken a comprehensive investigation of the magnetization behavior of the Cactus Graph Model (CGM), a quantum spin system defined on a cactus graph which admits an exact dimer singlet ground state, akin to the celebrated Shastry-Sutherland model. By investigating the singlet-triplet gaps, interactions between excited triplets, correlated hopping, and with the help of an effective hard-core boson representation, we shed light on the magnetization behaviors of CGM. We discover that owing to the geometry of the system single-triplet excitations are dispersionless. We unveil the emergence of density wave magnetization plateaus, and supersolid phases in the presence of an external magnetic field. Moreover, we highlight the key steps towards the formation of two- and potentially multi-particle extended bound states arising from correlated hopping. This could potentially lead to a further stabilization of density wave magnetization plateaus and the emergence of additional plateaus at lower magnetizations.

Further exploration is required to fully understand the magnetization behavior of CGM, especially in the vicinity of $m=0$. At present, our examination of two-particle repulsions only extends to third-order expansions. Nevertheless, extending this analysis to higher orders could potentially unveil more extensive repulsion effects between particles, particularly at longer ranges. One crucial investigation is centered around the crystallization phenomenon in scenarios involving two-particle, or even multi-particle, bound states under weaker magnetic fields. Additionally, the impact of three-body interactions at elevated magnetization levels remains an unexplored domain, warranting thorough investigation.

\textit{Acknowledgments.} The work is supported by the Deutsche Forschungsgemeinschaft (DFG, German Research Foundation) through Project-ID 258499086-SFB 1170 and the Würzburg-Dresden Cluster of Excellence on Complexity and Topology in Quantum Matter – ct.qmat Project-ID 390858490-EXC 2147.

%

\end{document}